\documentclass{pasj00}
\usepackage{psfig}
\begin{document}
\SetRunningHead{J. Yokogawa et al.}{X-Ray Supernova Remnants in the SMC}
\Received{2001/06/18}%{yyyy/mm/dd}
\Accepted{2001/12/27}%{yyyy/mm/dd}

\title{Centrally Peaked X-Ray Supernova Remnants in the 
Small Magellanic Cloud Studied with ASCA and ROSAT}
\author{%
Jun \textsc{Yokogawa},
Kensuke \textsc{Imanishi},
Katsuji \textsc{Koyama}}
\affil{Department of Physics, Graduate School of Science,
Kyoto University, Sakyo-ku, Kyoto 606-8502}
\email{jun@cr.scphys.kyoto-u.ac.jp}
\email{kensuke@cr.scphys.kyoto-u.ac.jp}
\email{koyama@cr.scphys.kyoto-u.ac.jp}
\author{%
Mamiko \textsc{Nishiuchi}}
\affil{Japan Atomic Energy Research Institute, 
Kansai Research Establishment, \\
8-1 Umebi-dai, Kizu-cho, Soraku-gun, Kyoto 619-0215}
\email{nishiuchi@apr.jaeri.go.jp}
\and
\author{%
Norikazu \textsc{Mizuno}}
\affil{Department of Astrophysics, Nagoya University, 
Furo-cho, Chikusa-ku, Nagoya 464-8602}
\email{norikazu@a.phys.nagoya-u.ac.jp}

\KeyWords{H~{\sc ii} regions --- ISM: abundances --- ISM: individual
(0103$-$726, 0045$-$734, 0057$-$7226) --- Magellanic Clouds ---
supernova remnants --- X-rays: ISM}
\maketitle

\begin{abstract}
This paper presents ASCA/SIS and ROSAT/HRI results of 
three supernova remnants (SNRs)
in the Small Magellanic Cloud: 0103$-$726, 0045$-$734, and 0057$-$7226. 
The ROSAT/HRI images of these SNRs 
indicate that the most of the X-ray emissions are  
concentrated in the center region. 
Only from 0103$-$726 are 
faint X-rays along the radio shell also detected. 
The ASCA/SIS spectra of 0103$-$726 and 0045$-$734
exhibit strong emission lines from highly ionized metals. 
The spectra were well-fitted with 
non-equilibrium ionization (NEI) plasma models. The
metal abundances are found to be larger than 
the mean chemical compositions in the interstellar medium (ISM) 
of the SMC. Thus, X-rays from these two SNRs
are attributable to the ejecta gas, although 
the ages estimated from the ionization timescale  are
significantly large, $\gtrsim 10^4$~yr.	
The chemical compositions are roughly consistent with the 
type-II supernova origin of a progenitor mass $\lesssim 20 M_{\solar}$.
The SIS spectrum of 0057$-$7226 was also fitted with an NEI model
of an estimated age $\gtrsim 6 \times 10^3$~yr.
Although no constraint on the metal abundances was obtained, 
the rather weak emission lines are  consistent with 
the low metal abundances in the ISM of the SMC. 
A possible scenario for the evolution  of the morphologies and spectra of SNRs 
is proposed. 
\end{abstract}

\section{Introduction}\label{sec:intro}

The Large and Small Magellanic Clouds (hereafter LMC and SMC)
are ideal galaxies  for systematic X-ray studies of SNRs,
because of their reasonable sizes, well-calibrated distances 
(50~kpc and 60~kpc, respectively; \cite{ber2000}), 
and low interstellar absorptions.

\citet{wil1999} obtained X-ray images of LMC SNRs 
mainly with ROSAT/HRI, and classified the SNRs into six 
classes according to the morphology. 
They found a loose correlation between the SNR sizes 
and the morphological classes: 
the smaller SNRs tend to exhibit shell-like X-ray emission, 
while the larger ones have ``diffuse-face'' or centrally-brightened
morphologies. 
This correlation indicates 
the evolution from shell-like to centrally-brightened SNRs. 
\citet{wil1999} argued that 
the centrally brightened morphologies could be produced 
by the cloud evaporation model \citep{whi1991} 
or the fossil radiation model  (\cite{rho1998}, and references 
therein). 

\citet{hug1998} systematically studied the ASCA/SIS spectra 
of seven middle-aged SNRs with a model assuming 
an internal structure based on the Sedov solution. 
They found that the average metal abundances in the SNRs decrease 
with increasing SNR age, and gradually approach to the mean value
of the interstellar medium (ISM) in the LMC.
\citet{nis2001} analyzed nine fainter LMC SNRs in a less-systematic way 
and determined the basic plasma parameters. 
Besides the general tendency of the abundance decrease, 
%overabundances for several elements were found from five SNRs 
%among the 16 SNRs examined by \citet{hug1998} and \citet{nis2001}. 
five SNRs among the 16 SNRs examined by \citet{hug1998} and \citet{nis2001} 
show overabundances of several elements 
(overabundant species are different for each SNR).

In the SMC, on the other hand, detailed X-ray studies have been
limited to the brightest and youngest SNR, 
0102$-$723, due mainly to the
far-fainter X-rays from the other SNRs. 
\citet{hay1994} analyzed the ASCA/SIS spectrum from the full region 
of 0102$-$723 and found overabundances of heavy elements, 
consistent with a type-II young SNR. 
With the Chandra ACIS, \citet{hug2000} have spatially resolved
X-rays from the blast-wave and those from the ejecta-dominated inner 
region; 
the former exhibits low elemental abundances consistent with the SMC ISM, 
while the latter is overabundant, especially in O and Ne.

The second-brightest X-ray SNR in the SMC 
is 0103$-$726, and the next luminous class includes 
several SNRs, such as  0045$-$734 and 0057$-$7226 \citep{hab2000}. 
All of these SNRs are located in H~{\sc ii} regions, 
DEM~S125, N19, and N66, respectively (e.g., \cite{fil1998}).
\citet{mil1982} carried out radio observations with 
the Molonglo Observatory synthesis telescope (MOST) 
and determined the diameters of the former 
two SNRs to be 
$\sim 52$~pc (0103$-$726) and $\sim 25$~pc (0045$-$734). 
The Einstein/HRI image of 0103$-$726 shows extended X-rays with the 
same size of radio emission \citep{ino1983}.

In order to distinguish SNRs from the strong background 
emission of the H~{\sc ii} region, \citet{ye1991} made a map of
the intensity difference between H$\alpha$ and the radio continuum,
 instead of the conventional method of using the intensity ratio of 
[S~{\sc ii}]/H$\alpha$. They then found a non-thermal shell structure 
with a diameter of  $\sim 56$~pc, the SNR 0057$-$7226, from
the most luminous H~{\sc ii} region in the SMC, N66. 
The sizes of these three SNRs are thus much larger than 
that of 0102$-$723, $\sim 8$~pc \citep{mil1982}.
By an analogy of the LMC samples used by \citet{hug1998}, 
these large sizes indicate that the three SNRs should be very old, 
possibly older than $10^4$~yr, or even more.
In fact, \citet{ros1994} carried out a  kinematic study in H$\alpha$ 
emission
and determined an age of 0045$-$734 to be $5.7 \times 10^4$~yr 
with a Sedov phase assumption. 
Usually, X-ray spectra from such large (old) SNRs are expected to exhibit 
low metal abundances consistent with the ISM, 
as has already been reported in the  LMC samples \citep{hug1998}.

However, a preliminary analysis of the ASCA spectra by \citet{yok2000} 
does not agree with the above prediction, 
except for 0057$-$7226 (N66): 
overabundances of several elements were suggested 
for 0103$-$726 and 0045$-$734 (N19). 
We therefore carried out more detailed and elaborate X-ray studies
on the three SNRs with a uniform analysis, 
using the high-resolution spectroscopy and imaging data 
of the ASCA/SIS and ROSAT/HRI, 
as well as information from radio observations.

\section{Observations and Data Reduction}\label{sec:obs}

ASCA observations of 0103$-$726, 0045$-$734, and 0057$-$7226 
were carried out 
with two GISs (Gas Imaging Spectrometers, \cite{oha1996}) and
two SISs (Solid-state Imaging Spectrometers, \cite{bur1994})
on the focal planes of the four XRTs (X-ray Telescopes, \cite{ser1995}). 
To study the spectra, we used only the SIS data, which
had better energy resolution ($\sim 2$\% at 5.9~keV) than the GIS;
hence, we do not refer to the GIS in this paper.
The date and SIS operation mode
of each observation are listed in table \ref{tab:obs}.

We rejected the SIS data obtained in the South Atlantic Anomaly,
in low cut-off rigidity regions ($<4$~GV),
when the elevation angle was low ($< 5^\circ$), 
or when the elevation from the bright earth was low ($< 25^\circ$). 
Hot and/or flickering pixels were also removed.
In order to restore  the quality degradation after
long-year operation in orbit, we applied a correction
of the Residual Dark Distribution (T.\ Dotani et al.\ 1997,
ASCA News 5, 14), which is available only for faint-mode data.
After these screenings, the total available exposure times were
$\sim 61$~ks for 0103$-$726, 
$\sim 134$~ks for 0045$-$734, and 
$\sim 34$~ks for 0057$-$7226 (see table \ref{tab:obs}).

Since the spatial resolution of ASCA is limited
($\sim 3'$ in a half-power diameter),
we used ROSAT/HRI data (resolution is $\sim 6''$) for an 
imaging study.
We retrieved screened image data of 
0103$-$726, 0045$-$734, and 0057$-$7226, 
for which the sequence IDs are 
rh500428, rh500136, and rh900445, respectively,  
from the HEASARC Online Service. 
The observation of each SNR was made by 
two or three separate exposures centered on the same coordinates. 
We combined the data from those exposures, 
resulting in total available exposure times of
$\sim 31$~ks for 0103$-$726, 
$\sim 12$~ks for 0045$-$734, 
and 
$\sim 49$~ks for 0057$-$7226 (see table \ref{tab:obs}).

\section{Analyses and Results}\label{sec:ana}
\subsection{ROSAT HRI Images}\label{subsec:image}

\begin{figure}[hbtp]
\hspace*{4mm}
\psfig{file=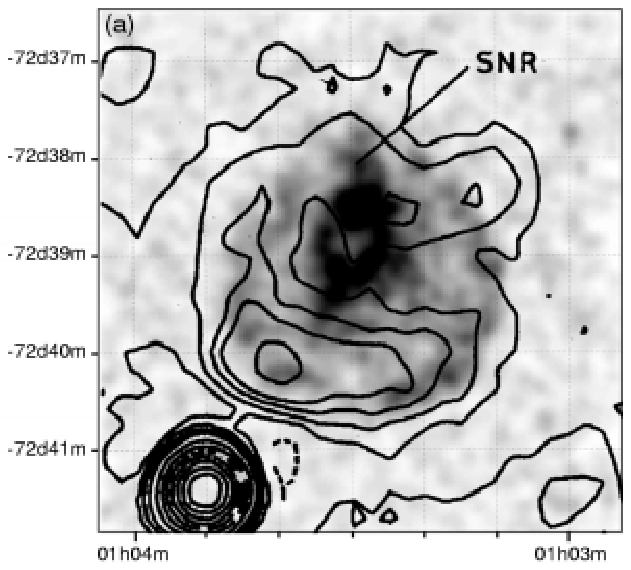,width=0.43\textwidth,clip=}

\hspace*{4mm}
\psfig{file=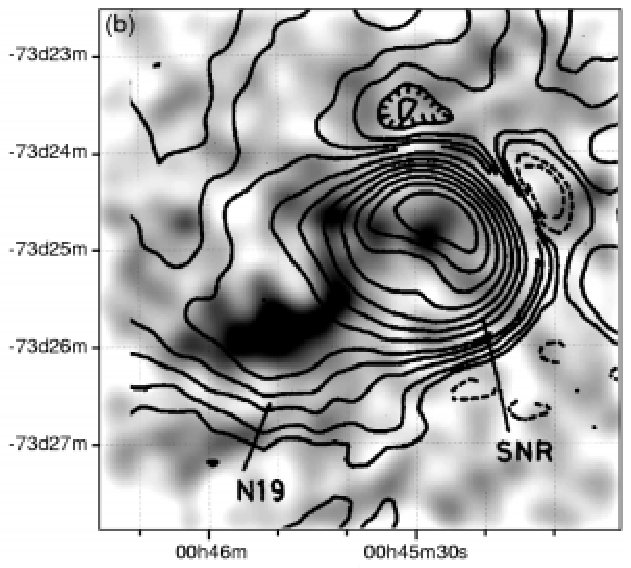,width=0.43\textwidth,clip=}

\hspace*{4mm}
\psfig{file=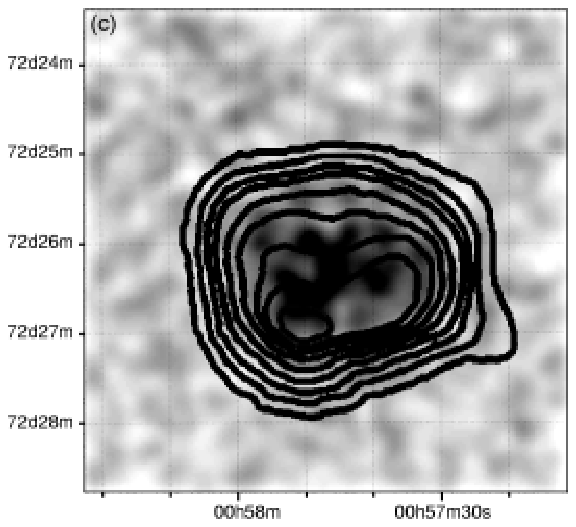,width=0.43\textwidth,clip=}

\caption{Gray-scale X-ray images of (a) 0103$-$726, (b) 0045$-$734, 
and (c) 0057$-$7226, 
obtained with ROSAT/HRI.
Overlaid contours are 
(a,b) the MOST radio image at 843~MHz taken from \citet{mil1982}, 
(c) the difference map of H$\alpha$ and MOST 843~MHz images 
reproduced from \citet{ye1991}. 
The equinox of the equatorial coordinates is 1950.0.
Strong radio emission at the southeast of 0103$-$726
is a background point source.
}
\label{fig:image}
\end{figure}

The X-ray images obtained with ROSAT/HRI are shown 
and compared with the radio continuum emission 
in figure \ref{fig:image}. 
Radio images at 843~MHz, obtained with MOST \citep{mil1982}, are overlaid 
for (a) 0103$-$726 and (b) 0045$-$734. 
For 0057$-$7226, the difference map of H$\alpha$ and 
the MOST images, which selectively reveals non-thermal radio emission 
\citep{ye1991}, are overlaid in (c). 

The radio emission of 0103$-$726 has a faint shell 
with an $\sim 3'$ diameter, 
in addition to enhanced emission at the southwest (SE) part.
The X-rays show a faint shell with an $\sim \timeform{2.5'}$ diameter 
along the radio shell,
which was already noted in an earlier result \citep{ino1983}.
More prominent X-rays are found near the center of the radio shell
with a small offset of $\sim 30''$ to the north.
These central X-rays exhibit a small elliptical ring.

The radio emission from 0045$-$734 is extended to
$\sim 3' \times 4'$, with a partial shell at the northwest (NW)
and a weak diffuse structure to the SE.
The NW shell exhibits strong [S~{\sc ii}] optical lines, while
several weak [S~{\sc ii}] filaments are found
within the SE emission \citep{mat1983}.
No shell-like structure is found in the X-ray band. 
Instead, the X-ray emission exhibits 
an irregular shape with 
strong emission at SE and faint knots at NW, 
both of which are 
confined to within the radio shell and diffuse structure.
According to \citet{dic2001}, the west-most knot may correspond to 
another SNR candidate, MCRX~J0047.2$-$7308. However, this knot is 
much fainter than the remaining part of the X-ray emission, 
and hence contamination from MCRX~J0047.2$-$7308 would be small. 
Therefore, the following results would fairly well represent the properties 
of 0045$-$734. 

The non-thermal radio emission from 0057$-$7226 
exhibits a shell-like structure with a diameter of $\sim 3'$, 
while the X-rays are concentrated in the radio shell, 
with a diameter of $\sim 2'$.

\subsection{ASCA SIS Spectra}\label{subsec:spec}

\begin{figure}[t]
\begin{center}
\hspace*{4mm}
\psfig{file=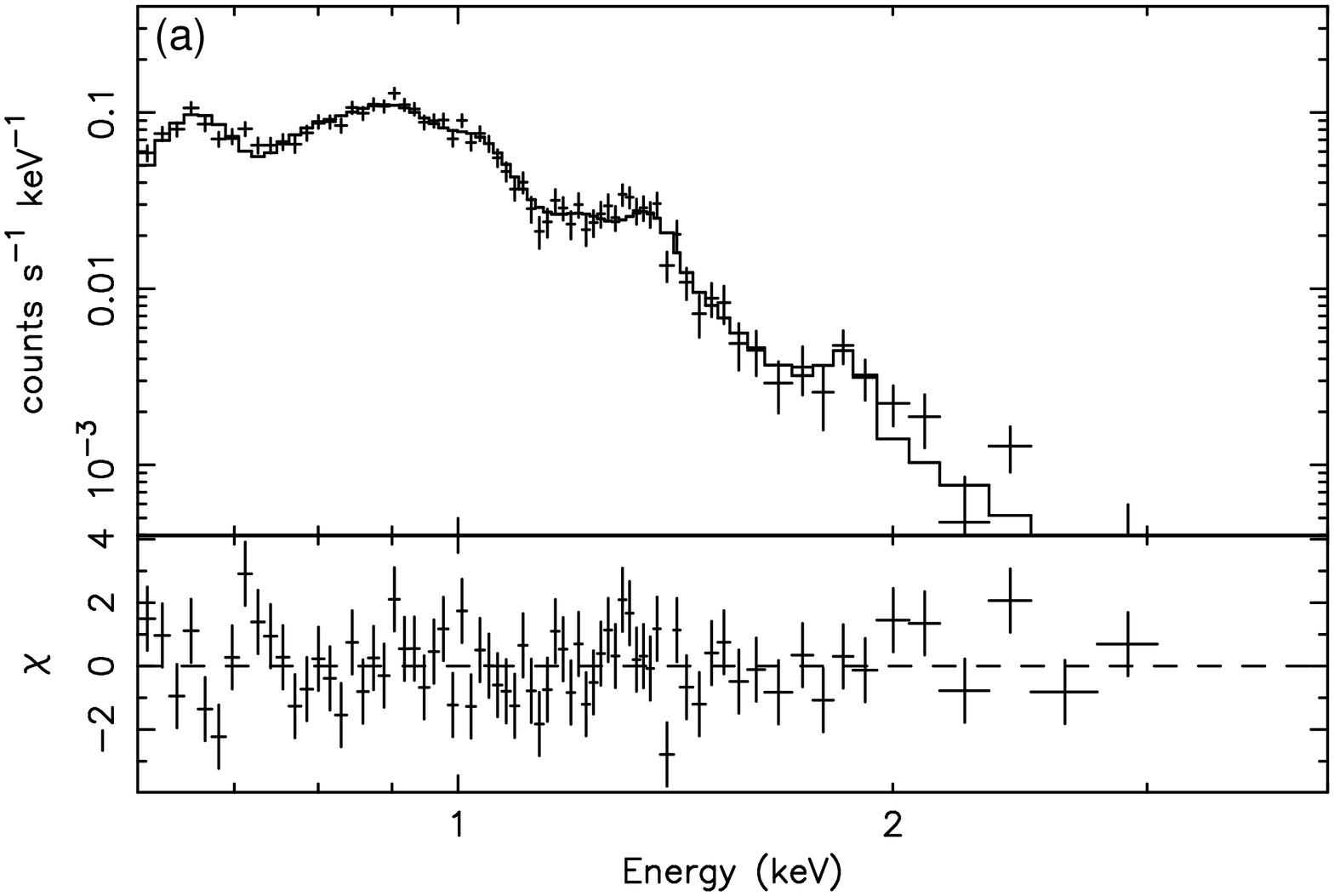,width=0.43\textwidth,clip=}

\hspace*{4mm}
\psfig{file=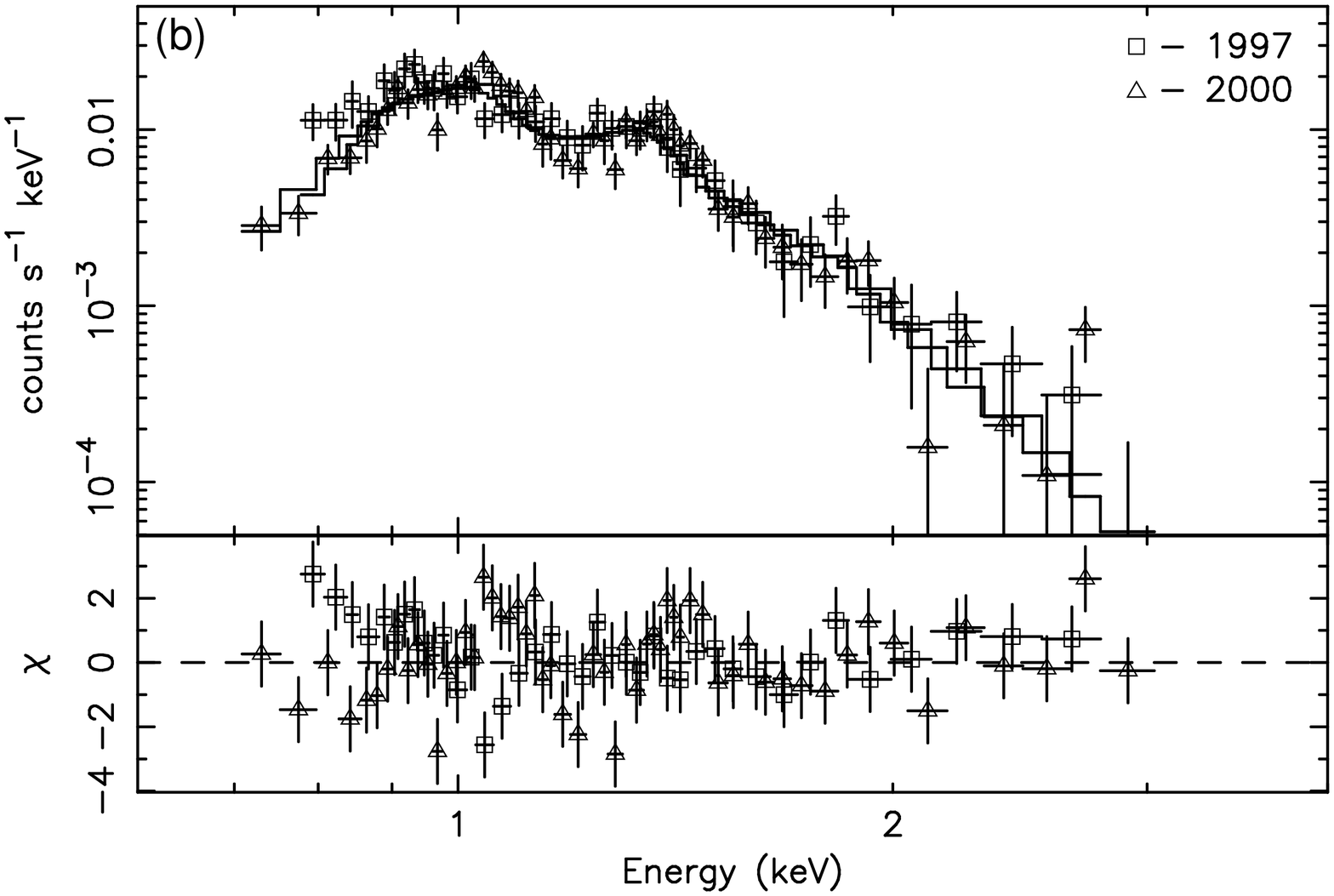,width=0.43\textwidth,clip=}

\hspace*{4mm}
\psfig{file=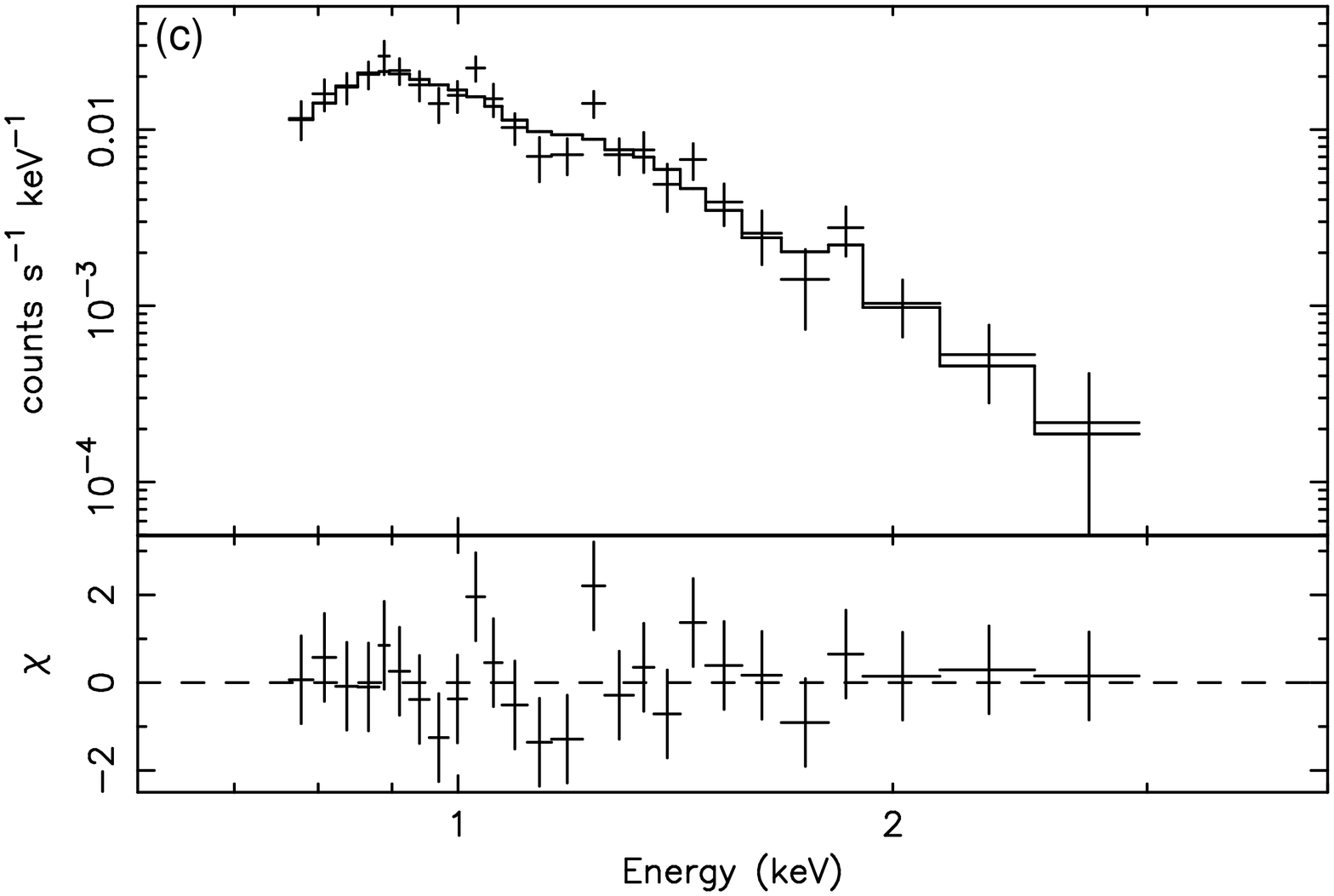,width=0.43\textwidth,clip=}
\caption{X-ray spectra of (a) 0103$-$726, (b) 0045$-$734, 
and (c) 0057$-$7226, 
obtained with ASCA SIS-0$+$SIS-1.
The crosses and the solid lines in each top panel indicate
the data points and the best-fit models
(model III for 0103$-$726 and model I for 0045$-$734;
see table \ref{tab:spec}), respectively.
The bottom panels show the residuals from the best-fit models.
The squares and triangles in (b) represent the observations in
1997 and 2000, respectively.
}
\label{fig:spec}
\end{center}
\end{figure}

We extracted the source spectra from the entire regions of each SNR,
while the background spectra were taken from nearby source-free 
regions.
The spectra from SIS-0 and SIS-1 were co-added so as to
increase the statistics.
The background-subtracted spectra are shown in figure
\ref{fig:spec}.
Since the spectra of these SNRs show emission lines 
from highly ionized heavy elements (\cite{yok2000}), 
we fitted the spectra with
an optically thin thermal-plasma model 
in a non-equilibrium ionization (NEI) state,
coded by \citet{mas1994}. The physical parameters
in the NEI model are the electron temperature ($kT_{\rm e}$),
the metal abundances, and the ionization timescale, 
$\tau =  n_{\rm e}t$, 
where $n_{\rm e}$ is the electron density and $t$ is the elapsed time
after the plasma was heated up.

We first fitted the spectrum of 0103$-$726 
with a fixed abundance of 0.2~solar, 
which is 
the mean value of the ISM of the SMC (\cite{rus1992}; 
hereafter, ``SMC mean value''). 
This model, however, left  bump-like (line-like) residuals 
at the energies of the K-shell emission lines from 
O, Ne, Mg, and Si, and Fe-L.  
Therefore, we allowed abundances of 
O, Mg, and Si to vary freely. 
In addition, 
since the Ne-K and Fe-L lines may equally contribute to 
X-ray emission around $\sim 1$~keV, 
we treated the abundances of Ne and Fe according to the following three cases: 
(I) Ne is free, (II) Fe is free, and (III) both Ne and Fe are free.
Model II, however, was rejected with 
%reduced $\chi^2 =  1.94$  for 64 degrees of freedom (d.o.f.), 
$\chi^2 = 124$ for 64 degrees of freedom (d.o.f.), 
while models I and III gave much better fits 
($\chi^2$ is 94.1 and 88.8, respectively). 
Even these two models were rejected from a statistical point of view,
possibly due to the residual uncertainty of the response function after 
the RDD correction. 
To extract a more accurate response is impractical at this moment; 
thus, we adopt these models in the following discussion as a reasonable
approximation. The best-fit parameters for models I and III are given in
table \ref{tab:spec}.
In both models, the heavy elements show larger abundances 
than the SMC mean value.

For 0045$-$734, the spectra from the two separate observations
were simultaneously fitted (all of the free parameters were linked between 
the two spectra).
We first fitted 
the spectra with an NEI model of fixed abundances to the 
SMC mean value. 
Since the residuals remained at energies of 
emission lines from Ne-K/Fe-L and Mg-K, 
we fit the spectra with an NEI model 
in which the abundance of Mg was a free parameter, 
and those of Ne and Fe were treated as described above (models I--III). 
For model III, the best-fit temperature is extremely large, 
$kT_{\rm e}\sim 25$~keV. 
Such a high temperature would be unphysical 
because youngest SNRs ever known exhibit a much lower temperature 
(several keV). 
We thus do not adopt model III. 
%For model III, we cannot get a good fit within a reasonable
%temperature range, thus do not adopt model III 
%(the best-fit temperature is extremely large, $kT_{\rm e}\sim 25$~keV).
Models I and II, on the other hand, gave equally better fits  with 
no bump-like (line-like) residuals.
For the same reason described above, 
we adopt these models hereafter. 
The best-fit parameters of models I and II are given in
table \ref{tab:spec}.
In both models, the abundances of Mg and Ne/Fe are  larger than
those of the SMC mean value.

Although the spectrum of 0057$-$7226 was well-fitted with 
an NEI model with free-abundances, 
we could not constrain the abundances due to the limited statistics. 
We hence fixed the abundances to be 0.2~solar 
(the SMC mean value). 
The best-fit parameters are given in 
table \ref{tab:spec}. 

\section{Discussion}\label{sec:dis}
\subsection{Plasma Parameters in the SNRs}\label{subsec:plasma}

Using the X-ray images (figure \ref{fig:image}), 
we estimated the volume $V$ of the X-ray emitting plasma. 
For simplicity, we assumed that the plasma in 0103$-$726 and 0057$-$7226 
fills a sphere with a diameter of \timeform{2.5'} and $2'$, respectively,  
while that in 0045$-$734 fills 
a cylinder with a height of $3'$ and a diameter of $1'$. 
We then obtained 
$V =  1.3 \times 10^{60} \cdot \beta$~cm$^{3}$ for 0103$-$726,
$V =  3.6 \times 10^{59} \cdot \beta$~cm$^{3}$ for 0045$-$734, and
$V =  6.6 \times 10^{59} \cdot \beta$~cm$^{3}$ for 0057$-$7226, 
respectively, where 
$\beta$ represents the volume filling factor, 
which could be different for each SNR.

From  the emission measure $EM$ determined by the spectral fitting, 
we could estimate the electron density of the plasma, $n_{\rm e}$, 
by $n_{\rm e} =  \sqrt{EM/V}$. 
The number density of the nucleons was simply assumed 
to be the same as that of electrons. 
The age $t$ was then determined from the ionization timescale, 
$\tau$, by $t =  \tau/n_{\rm e}$.
The total mass of the plasma $M_{\rm total}$ was
estimated by  $M_{\rm total} =  n_{\rm e}Vm_{\rm H}$,
where $m_{\rm H}$ is the mass of a hydrogen atom. 
We then determined the mass of the overabundant heavy elements, 
using $M_{\rm total}$ and the best-fit abundances.
These estimated parameters are given in table 
\ref{tab:plasma}. 
The lower limit for the ionization age of 0045$-$734 
is  $2.9\times 10^4$~yr in both models, 
which is consistent with the age of $5.7 \times 10^4$~yr derived from 
an optical study 
of the kinematics \citep{ros1994}. 

\begin{figure}[hbtp]
\begin{center}
\hspace*{4mm}
\psfig{file=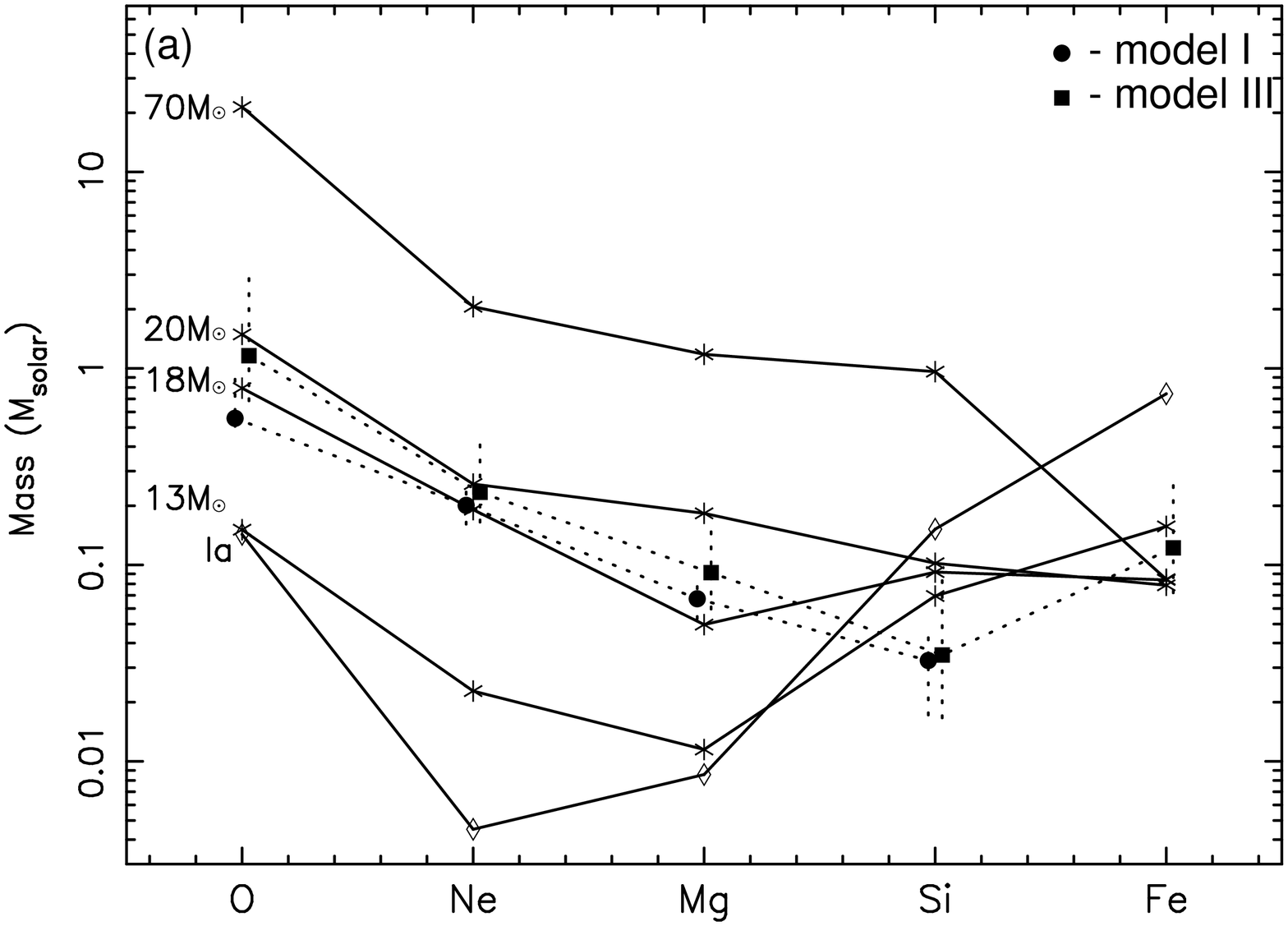,width=0.46\textwidth,clip=}

\hspace*{4mm}
\psfig{file=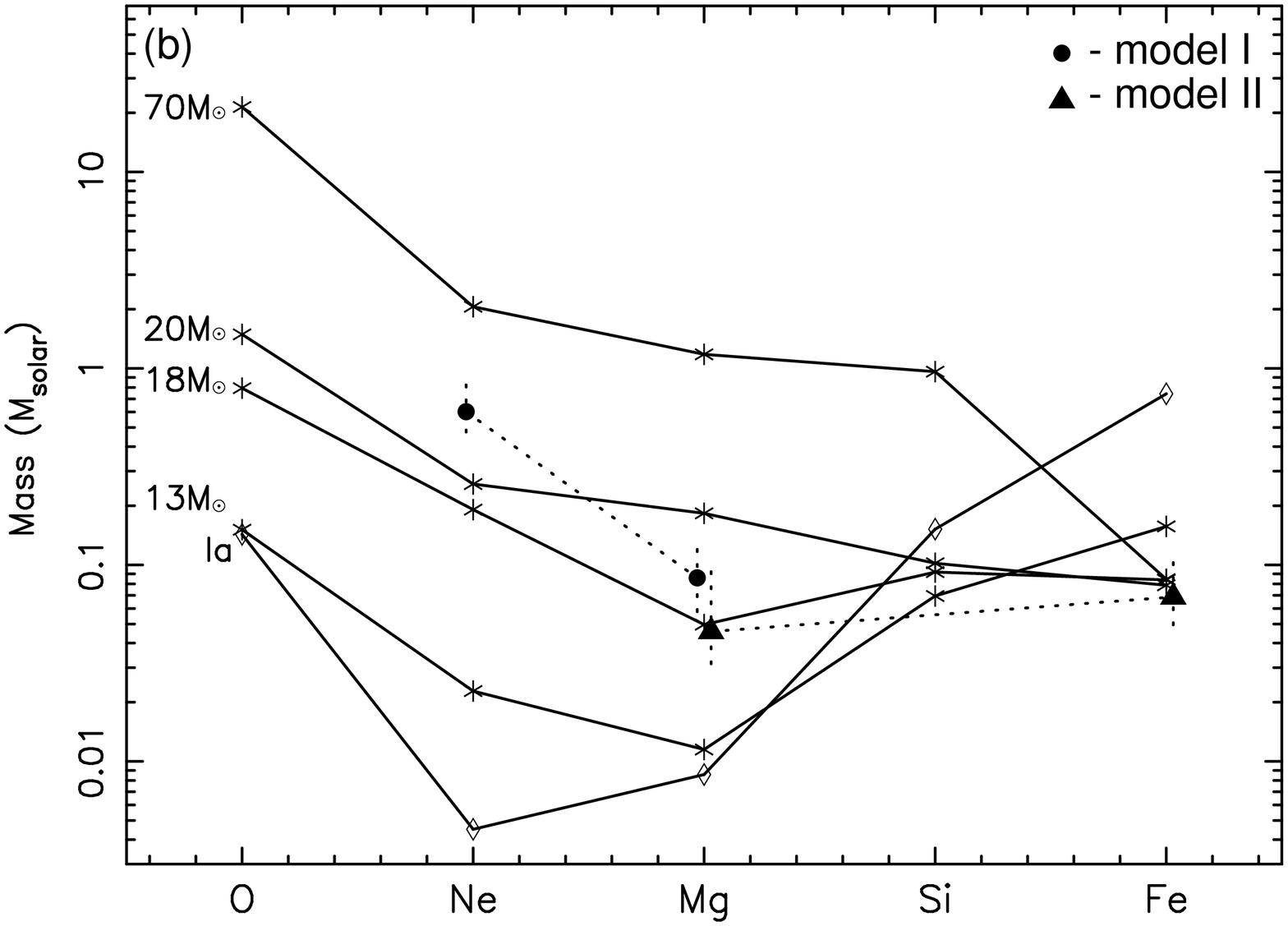,width=0.46\textwidth,clip=}
\caption{Nucleosynthesis products of various SNe
overlaid with the masses of elements in (a) 0103$-$726 and 
(b) 0045$-$734
derived from the spectral analyses.
Solid lines with asterisks and diamonds represent
products in type-II and type-Ia SNe, respectively
\citep{tsu1995}.
The progenitor masses of the type-II SNe are also indicated.
The dotted lines with circles, triangles, and squares
represent models I, II, and III, respectively.
Model II for 0103$-$726 and model III for 0045$-$734 
are not plotted because of the bad fitting (see subsection 
\ref{subsec:spec}).
Plots for the elements fixed to be 0.2 solar in the fitting
are not presented.
}
\label{fig:yield}
\end{center}
\end{figure}

For 0103$-$726 and 0045$-$734, 
the abundances are significantly larger than
the SMC mean value in any model; hence, the ejecta should largely contribute 
to the X-ray emitting plasma. 
In figure \ref{fig:yield}, we compare 
the estimated element masses for 0103$-$726 and 0045$-$734 
with 
the theoretical predictions of chemical compositions 
produced with supernovae (SNe) from various progenitor masses 
\citep{tsu1995}. 
Type-Ia origin would be rejected for both of the SNRs, 
because the data largely exceed the theoretical prediction for Ne and Mg, 
while the data do not show a large excess of Fe, as expected from theory.
%because the data
%do not show large excess of Fe as expected from the theory.
On the other hand, our data roughly agree with the prediction for
a type-II SN case with a progenitor mass of $\lesssim 20 M_{\solar}$.
Since these SNRs are located in H~{\sc ii} regions, the type-II SNe origin
is reasonable.

The spectrum of 0057$-$7226 does not exhibit overabundances, 
and is consistent with the SMC ISM. Therefore, 
the type of the SN cannot be constrained. 
However, its location in a giant H~{\sc ii} region
prefers a type-II origin.

Since the total mass of the ``metal-poor'' SNR 0057$-$7226
is comparable to those of ``metal-rich'' SNRs 0103$-$726
and  0045$-$734,  0057$-$7226 may simply have 
a smaller ejecta mass than the other two SNRs. 
One may argue, alternatively, if the progenitor is 
less massive than those of 0103$-$726 and 0045$-$734,
for example   $\sim 13 M_{\solar}$, that the abundances of
O, Ne, and Mg may be explained (see solid lines in figure \ref{fig:yield}). 
However, this model predicts a larger mass of  Fe,
which is not found in the observed spectrum.

\subsection{Implication for the SNR Evolution}

We have shown that the X-ray emissions of the three SNRs 
are predominantly concentrated inside the radio shells 
(hereafter, ``centrally peaked'' morphology). 
The X-ray spectra predict that the ages are all  
rather old ($\gtrsim 10^4$~yr), 
while two of them are found to exhibit high abundances possibly due 
to the ejecta gas. 

To explain these features 
as well as the results from the LMC SNRs 
(\cite{wil1999}; \cite{hug1998}; \cite{nis2001}), 
we  propose the following scenario. 
Soon after an SNR enters the Sedov phase 
(i.e., middle-aged stage), 
X-ray emission from the swept-up ISM becomes dominant 
because of the higher density at the shell, 
and hence the SNR exhibits a shell-like morphology and a 
low-abundance spectrum. 
As the age increases, the abundances gradually decrease 
and approach the mean value of the ISM, 
due to dilution with the swept-up ISM. 
At the same time, the morphology gradually changes from shell-like 
(``Shell'' and ``Diffuse Face 1'' in \cite{wil1999})
to centrally-peaked type 
(``Diffuse Face 2'' and ``Centrally Brightened'' in \cite{wil1999}), 
due to rapid cooling of the shell compared with the inner region.
X-rays from the central region may thus be fossil radiation 
(\cite{rho1998}, and references therein) and could be enhanced by  
the evaporation of cloudlets \citep{whi1991}. 
The X-ray-emitting plasma shows a scatter of metal abundances, 
depending  on the  ejecta mass (large ejecta mass for  0103$-$726 and 0045$-$734
and small ejecta mass for  0057$-$7226). 
This scenario would be supported by 
numerical simulations by \citet{she1999}. 

The middle-aged LMC SNRs analyzed by \citet{hug1998} and \citet{nis2001} 
contain five shell-like and  seven centrally-peaked SNRs \citep{wil1999}; 
among them, one shell-like and three centrally-peaked SNRs 
were found to be overabundant. 
The fact that the ratio of the overabundant SNRs 
is higher for the centrally-peaked (3/7) than for the shell-like (1/5) 
may support the above scenario, 
although the statistics are still limited. 

In a transition phase from a shell-like to a centrally-peaked SNR, 
a faint X-ray shell, like that of 0103$-$726, may be observed.
In this case, the X-ray spectrum of the faint shell should exhibit 
low abundances consistent with the SMC ISM, 
while the inner part should exhibit 
an ejecta-dominated spectrum. 
Spatially resolved spectroscopy with Chandra or XMM-Newton 
should clarify this prediction. 
\par
\vspace{1pc}\par
The authors are grateful to Dr.\ K.\ Yoshita, 
whose strong criticism and useful comments made the initial draft much better. 
The authors also thank constructive comments from 
the referee, Dr.\ R.\ Williams. 
J.Y.\ and K.I.\ are
supported by the JSPS Research Fellowship for Young Scientists.
We retrieved ROSAT data from the HEASARC Online System
which is provided by NASA/GSFC.

\onecolumn

\begin{table}
\caption{Observation log.}\label{tab:obs}
\begin{center}
\begin{tabular}{llcl}
\hline\hline
\multicolumn{4}{c}{ASCA/SIS}\\
\multicolumn{1}{c}{Target} &
\multicolumn{1}{c}{Date}   & 
Exposure (ks)&
\multicolumn{1}{c}{SIS mode$^\ast$}    \\
\hline              
0103$-$726  &1996/05/21--23& 60.8         &1-CCD Faint/Faint  \\
0045$-$734  &1997/11/13--14& 37.8         &2-CCD Faint/Faint  \\
0057$-$7226 &1997/11/14--15& 33.8         &2-CCD Faint/Faint  \\
0045$-$734  &2000/04/11--17& 95.9         &2-CCD Faint/Bright \\
\hline\hline
\multicolumn{4}{c}{ROSAT/HRI}\\
\multicolumn{1}{c}{Target} &
\multicolumn{1}{c}{Date}   & 
Exposure (ks)& \\
\hline              
0045$-$734  &1994/04/17--18   &\phantom{1}4.4          & \\
0057$-$7226 &1994/04/19--06/09&14.8                    & \\
0045$-$734  &1994/04/21--05/16&\phantom{1}7.4          & \\
0057$-$7226 &1995/04/13--05/12&34.3                    & \\
0103$-$726  &1996/12/13       &\phantom{1}1.1          &  \\
0103$-$726  &1997/03/28--04/14& 16.7          &  \\
0103$-$726  &1997/04/19--06/12& 12.8          &\\
\hline
\multicolumn{3}{l}
{$^{\ast}$ Data format in the high/medium bit rate.}
\end{tabular}
\end{center}
\end{table}

\begin{table}
\begin{center}
\caption{Best-fit parameters for the spectrum of the SNRs.}\label{tab:spec}
\begin{tabular}{llllll}
\hline\hline
SNR &\multicolumn{2}{c}{0103$-$726}&
 \multicolumn{2}{c}{0045$-$734}&
 \multicolumn{1}{c}{0057$-$7226}\\
Model$^*$ & \multicolumn{1}{c}{I} & \multicolumn{1}{c}{III} 
& \multicolumn{1}{c}{I} & \multicolumn{1}{c}{II} 
& \multicolumn{1}{c}{...} \\
\hline
$\log{\tau}$ (s cm$^{-3}$) &
 10.97$^{+0.21}_{-0.21}$& 10.82$^{+0.25}_{-0.26}$ &
 13.00$^\S${}$^{+\infty}_{-1.20}$& 11.74$^{+\infty}_{-0.41}$ &
 11.01$^{+\infty}_{-0.44}$\\
$kT_{\rm e}$ (keV) &
 0.67$^{+0.13}_{-0.10}$ & 1.0$^{+0.5}_{-0.2}$ &
 0.37$^{+0.10}_{-0.05}$ & 0.96$^{+0.08}_{-0.05}$ &
 0.9$^{+0.8}_{-0.3}$ \\
$N_{\rm H}$ ($10^{21}$ H cm$^{-2}$) &
  0$^{+0.5}$ & 0$^{+0.3}$  &
 10$^{+1}_{-2}$ & 3$^{+1}_{-2}$  &
 1$^{+2}_{-1}$ \\
Norm$^\dagger$ ($10^{11}$~cm$^{-5}$) &
 0.86& 0.28&
 3.9 & 0.45&
 0.51\\
Flux$^\ddagger$ (erg s$^{-1}$ cm$^{-2}$) &
 $1.1 \times 10^{-12}$ & $1.0 \times 10^{-12}$ &
 $2.9 \times 10^{-13}$ & $3.1 \times 10^{-13}$ &
 $3.0 \times 10^{-13}$ \\
--- Abundances (solar unit) --- &&\\
\ \ O &
  0.5$^{+0.3}_{-0.1}$ & 1.7$^{+3.0}_{-0.7}$&
  0.2 (fixed) & 0.2 (fixed)&
  0.2 (fixed) \\
\ \ Ne &
  1.0$^{+0.2}_{-0.2}$ & 1.9$^{+1.7}_{-0.6}$&
  2.6$^{+1.0}_{-0.7}$ & 0.2 (fixed)&
  0.2 (fixed) \\
\ \ Mg &
  0.9$^{+0.3}_{-0.2}$ & 2.0$^{+1.8}_{-0.8}$ &
  1.0$^{+0.5}_{-0.4}$ & 1.6$^{+1.7}_{-0.6}$ &
  0.2 (fixed) \\
\ \ Si &
  0.4$^{+0.2}_{-0.2}$ & 0.7$^{+1.1}_{-0.4}$ &
  0.2 (fixed) & 0.2 (fixed)&
  0.2 (fixed) \\
\ \ Fe &
  0.2 (fixed)    & 1.0$^{+1.1}_{-0.5}$ &
  0.2 (fixed)    & 0.9$^{+0.5}_{-0.3}$ &
  0.2 (fixed) \\
Reduced $\chi^2$/d.o.f. &
  1.47/64  & 1.41/63  &
  1.45/89  & 1.60/89  &
  0.91/22 \\
\hline
\multicolumn{6}{l}{Note. Errors indicate the 90\% confidence 
limits.}\\
\multicolumn{6}{l}{$^*$ Defined in subsection \ref{subsec:spec}.}\\
\multicolumn{6}{l}{$^\dagger$ Norm $\equiv EM/4\pi D^2$, where $D$ is the 
distance (60~kpc) and $EM$ is the emission measure. }\\
% norm = 1e-12*n_i*n_i*V/4piD^2 = 1e-12*EM/4piD^2 in cgs (masai89)
\multicolumn{6}{l}{$^\ddagger$ In 0.7--10.0~keV.}\\
\multicolumn{6}{l}{$^\S$ Pegged by the upper limit of the plasma 
code.}\\
\end{tabular}
\end{center}
\end{table}

\begin{table}
\caption{Parameters derived from the best-fit models for the SNRs.}\label{tab:plasma}
\begin{center}
\begin{tabular}{lccccc}
\hline\hline
SNR &\multicolumn{2}{c}{0103$-$726}&
 \multicolumn{2}{c}{0045$-$734}&
 \multicolumn{1}{c}{0057$-$7226}\\
Model$^*$ & I & III & I & II & ... \\
\hline
$V$ ($\beta$ cm$^{3}$) &
\multicolumn{2}{c}{--- $1.3 \times 10^{60}$ ---}&
\multicolumn{2}{c}{--- $3.6 \times 10^{59}$ ---}&
$6.6 \times 10^{59}$ \\
$n_{\rm e}$ ($\beta^{-1/2}$ cm$^{-3}$) &
0.17&0.097&
0.69&0.23&
0.18\\
$t$ ($\beta^{1/2}$ yr) &
1.8$^{+1.0}_{-0.7}  \times 10^4$&2.2$^{+1.6}_{-1.0} \times 10^4$&
($> 2.9  \times 10^4$)&7.6$^{+\infty}_{-4.7} \times 10^4$&
1.8$^{+\infty}_{-1.2}  \times 10^4$\\
$M_{\rm total}$ ($\beta^{1/2} M_{\solar}$) &
180&110&
210& 69&
100\\
Element mass ($\beta^{1/2} M_{\solar}$) &&\\
\ \ O &
5.6$^{+3.3}_{-1.1} \times 10^{-1}$&1.2$^{+2.0}_{-0.5}$ &
...$^\dagger$&...$^\dagger$ &
...$^\dagger$\\
\ \ Ne &
2.0$^{+0.4}_{-0.4} \times 10^{-1}$&2.3$^{+2.1}_{-0.7} \times 10^{-1}$&
6.0$^{+2.3}_{-1.6} \times 10^{-1}$&...$^\dagger$&
...$^\dagger$\\
\ \ Mg &
6.7$^{+2.3}_{-1.5} \times 10^{-2}$&9.1$^{+7.9}_{-3.6} \times 10^{-2}$&
8.6$^{+4.4}_{-3.4} \times 10^{-2}$&4.6$^{+4.9}_{-1.7} \times 10^{-2}$&
...$^\dagger$\\
\ \ Si &
3.3$^{+1.6}_{-1.7} \times 10^{-2}$&3.5$^{+5.5}_{-2.0} \times 10^{-2}$&
...$^\dagger$&...$^\dagger$ &
...$^\dagger$\\
\ \ Fe &
...$^\dagger$                  &1.2$^{+1.4}_{-0.6} \times 10^{-1}$&
...$^\dagger$                  &6.9$^{+4.1}_{-2.3} \times 10^{-2}$&
...$^\dagger$\\
\hline
\multicolumn{6}{p{\textwidth}}{Note. The volume filling factor is
represented by $\beta$. 
Errors are derived from the 90\% confidence limits for 
$\tau$ and elemental abundances in the spectral fitting 
(table \ref{tab:spec}).}\\
\multicolumn{6}{l}{$^*$ Defined in subsection \ref{subsec:spec}.}\\
\multicolumn{6}{l}{$^\dagger$ Fixed to be 0.2 solar abundance in the 
spectral fitting.}\\
\end{tabular}
\end{center}
\end{table}

\end{document}